# Effects of electrostatic force on piezoelectric materials under high electric field: Impact on GaN-based nanoscale structures

Carlos Rivera[1,2]

[1]*Ingeniería y Servicios Aeroespaciales S.A., Paseo Pintor Rosales 34, 28008 Madrid, Spain*
[2]*Unidad de Sensores, Micro y Nanotecnología, Instituto Tecnológico La Marañosa, San Martín de la Vega, 28330 Madrid, Spain*

E-mail: carlos.rivera@insa.es

**Abstract.** The determination of the electromechanical properties of materials for a parallel-plate capacitor structure is affected by the electrostatic force between their electrodes. The corrections induced by this electric-field-induced stress are usually assumed to be linked to the quadratic dependence of the strain on the electric field (electrostriction). Here we show by calculations based on thermodynamic grounds for this simple structure that the effect of the electrostatic force on piezoelectric materials can lead to both quadratic and linear corrections through the combination of the piezoelectric coupling and spontaneous polarization. The case of GaN-based capacitor nanoscale structures is presented taking into account the boundary conditions imposed as well as the effect of geometry. The results in this example point to corrections in the piezoelectric and electrostrictive coefficients higher than 0.3 pmV$^{-1}$ and 2.6×10$^{-22}$ m$^2$V$^{-2}$, respectively.

**1. Introduction**

The size reduction at the nanoscale has made possible to extend the capabilities of current devices and structures beyond the advantages expected due to the changes in geometry, as well as to observe new physical phenomena (e.g., see Refs. [1−7]). In particular, the interest in studying the fundamental properties of piezoelectric materials has grown as a consequence of the development of new nanotechnology-based concepts [8−11]. Recent experimental evidences point to a variation in the piezoelectric response for nanometric structures, such as carbon nanotubes or ZnO nanobelts [12,13]. However, the mechanisms affecting piezoelectricity at the mesoscopic and nanoscopic scales are still unclear. One possible explanation would lie in the increase of the Coulomb interaction ($\propto 1/L^2$ law, being $L$ the distance) between boundary or surface charges of external origin (e.g., at the electrodes of a capacitor structure) as the size of the structure shrinks. This electrostatic force due to the boundary charge has been usually neglected in the theoretical treatment of piezoelectric structures and devices since it is expected to play a minor role on their electromechanical properties, except in the case of some electro-active elastomers and polymers [14,15].

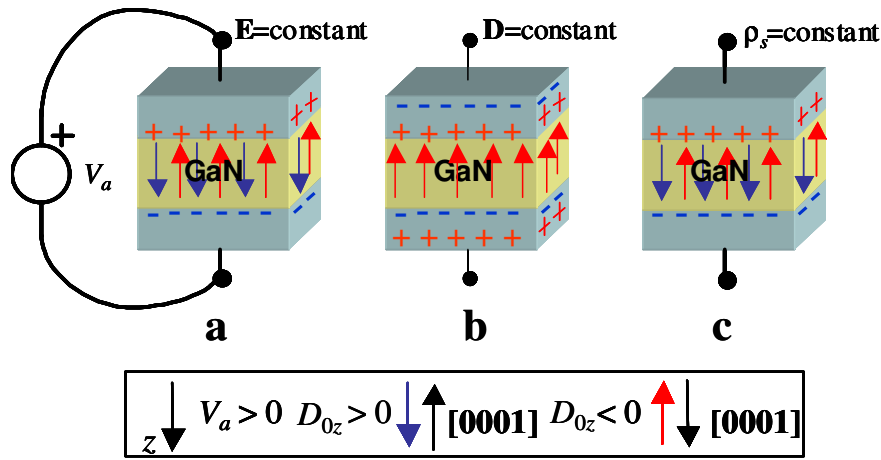

**Figure 1.** Schematic view of the studied structure for different boundary conditions. (a) the GaN-slab is sandwiched between two electrodes connected to a voltage source. (b) the GaN-slab is sandwiched between two electrodes in open circuit conditions. (c) the GaN-slab is sandwiched between two charged electrodes. The sign convention with respect to the [0001] direction of GaN is shown in the legend for all cases. Red up arrows apply for positive bias, whereas blue down arrows apply for negative bias [see panels (a) and (c)]. Thus, the sign of the spontaneous polarization ($D_{0z}$) with respect to $z$ is negative for $V_a > 0$ and positive for $V_a < 0$ (note that $z$ is opposite to [0001] in the latter case), whereas $V_a > 0$ for $z$ pointing down in the figure.

In addition to the practical interest in understanding the experimental results at the nanoscale, the study of the electrically-induced stress on piezoelectric materials is a subject of basic significance. The classical approach to address this phenomenon involves the concepts of electrostriction and Maxwell stresses [16,17]. Further works have reported the action of the electrostatic field on the electrostrictive response including the dependence of permittivity on strain [18,19]. However, the analysis of piezoelectric materials subjected to similar electrostatic conditions is lacking. The results presented here show that qualitative changes can appear in this particular case, affecting both the piezoelectric and electrostrictive responses. The linear dependence related to the change in the piezoelectric coefficient is found to be linked to the spontaneous polarization through the piezoelectric response to the electrically-induced stress. For the purposes of this article, the electrically-induced stress corresponds to the electrostatic-induced pressure (hereafter electrostatic force) exerted by the boundary charge on the strain. Note that, in this context, the boundary charge stands for the free surface charge on the electrodes arising from the boundary conditions of the electrostatic problem.

The effect of boundary conditions on the electrically-induced changes is also discussed. To that end, three alternative actuation schemes (constant-electric field, constant-electric displacement and constant-surface charge density) were applied to the same basic structure. The calculations were performed for the simple situation where the electric field is uniform for a parallel-plate capacitor GaN-based structure. Although, the quantitative results are restricted to the studied case, the method and main conclusions can be readily extended to other semiconductors such as ZnO.

## 2. Theoretical method

The structure used for the calculations consists of a piezoelectric GaN slab sandwiched between two metallic electrodes, i.e., the typical parallel-plate capacitor (see Fig. 1). The layers were assumed to be of infinite extent in the in-plane directions throughout the entire analysis to allow ignoring the effect of the fringing fields at the edges. The cases shown in Fig. 1 correspond to the different boundary conditions imposed on the electric field **E**, the electric displacement **D**, and the surface charge density $\rho_s$. The main difference arising from these cases is the one related to the value of the depolarizing field and, consequently, the surface charge density. The constant-**E** condition appears when a voltage generator is connected to the sample. In this case, the piezoelectric material must be considered as a good insulator, i.e. virtually depleted of free carriers. Here, the charge at the boundaries reflects the value of the electric field, where the voltage generator acts as a reservoir of charge to maintain the value of the applied voltage. The constant-**D** condition is appropriate to represent the situation where the electrodes are neutral [e.g., an open-circuit arrangement as the one depicted in Fig. 1(b)], giving rise to a different value for the depolarizing field. Alternatively, the constant-$\rho_s$ condition, whose practical implementation could involve a current or charge source, resembles the constant-**E** condition, except that the net charge on the electrodes is fixed. Hence, both **E** and the strain tensor **u** must be determined. In general, the consequences of using a particular actuation scheme are significant in terms of the equilibrium (or stationary) conditions reached, and also on the results of calculations [20,21]. Nevertheless, we will only focus on the constant-**E** condition, since our calculations show that the effects are smaller for the constant-$\rho_s$ condition and vanish for the constant-**D** condition.

The calculations were performed by minimizing the electric enthalpy functional subjected to the proper boundary conditions to obtain **E** and **u** in the GaN slab. The use of this method allows covering a broad range of thicknesses for the GaN slab, denoted as $d$ in equations, from the macroscopic to mesoscopic scales by means of the same definition for the strain. Furthermore, the theory of continuum elasticity is well behaved for small deformations up to a few interatomic distances. Thus, the electric enthalpy functional is written as

$$\Im_E = \frac{1}{2} C_{ijkl} u_{ij} u_{kl} - D_{0i} E_i - \frac{1}{2} \varepsilon_{ij} E_i E_j - e_{ijk} E_i u_{jk}, \quad (1)$$

where $C_{ijkl}$ is the fourth-ranked elastic stiffness tensor, $\varepsilon_{ij}$ the electric permittivity tensor, and $e_{ijk}$ the piezoelectric coefficient tensor. The term $D_{0i}$ denotes the spontaneous polarization, which only exists along the $[0001]$ direction in wurtzite materials. The boundary conditions were implemented including the stress produced by the electrostatic force between the electrodes. Taking into account that the general form for the boundary conditions at the surface is given by $\sigma_{ij} \hat{n}_j = f_i$, where $\sigma_{ij}$ is the stress tensor, $\hat{\mathbf{n}}$ is a vector normal to the surface and **f** is the force applied to the surface, and since the electrostatic pressure can be calculated as

$$\mathbf{P_{elec}} = \frac{1}{A} \iint \rho_s \mathbf{E} dS, \quad (2)$$

with $A$ being the area of the surface, the constant-**E** condition at an applied voltage $V_a$, as shown in Fig. 1(a), leads to the following boundary conditions

$$\sigma_{zz} = -\frac{D_z^2}{\varepsilon_{zz}}, \quad (3)$$

$$V_a = E_z d(1 + u_{zz}). \quad (4)$$

Very large values of $u_{zz}$ in Eq. (4) could initiate an instable behaviour in the response of the system to the electrostatic actuation. However, the correction introduced in Eq. (4) for $d$ due to $u_{zz}$ can be in general neglected for strain values below 5%.

## 3. Results and discussion

The analytical solution of the $\Im_E$ minimization problem subjected to the nonlinear constraints given by Eqs. (3) and (4) can be expressed in the form

$$\alpha_4 u_{zz}^4 + \alpha_3 u_{zz}^3 + \alpha_2 u_{zz}^2 + \alpha_1 u_{zz} + \alpha_0 = 0, \qquad (5)$$

$$u_{xx} = u_{yy} = \frac{\beta_2 u_{zz}^2 + \beta_1 u_{zz} + \beta_0}{\gamma_1 u_{zz} + \gamma_0}, \qquad (6)$$

$$u_{xy} = u_{yz} = u_{zx} = E_x = E_y = 0, \qquad (7)$$

for isotropic $\varepsilon_{ij}$ (i.e., $\varepsilon_{ij} = \varepsilon \ \forall i, j$) and neglecting the correction in Eq. (4) due to $u_{zz}$, where $\alpha_i$, $\beta_j$ and $\gamma_k$ are the coefficients including the $E_z$ dependence with $V_a$, as well as the various elastic, electric and piezoelectric parameters (see appendix for the full expressions). The solution derived using this method is restricted to quasi-static equilibrium conditions, in this case, justified for electrostatic actuation. Another solution of practical interest can be obtained imposing additional free surface boundary conditions ($\sigma_{xx} = \sigma_{yy} = 0$). The number of degrees of freedom is reduced adding these constraints, so that Eqs. (5) and (6) become $\alpha'_2 u'^2_{zz} + \alpha'_1 u'_{zz} + \alpha'_0 = 0$ and $u'_{xx} = (-C_{13} u'_{zz} + e_{31} E_z)/(C_{11} + C_{12})$, respectively (see also appendix for the full expressions). To carry out the analysis, we first study separately the contributions of the electrostatic force and piezoelectric effects to **u**, so that we can determine their relative importance. Subsequently, we perform the fully coupled calculations.

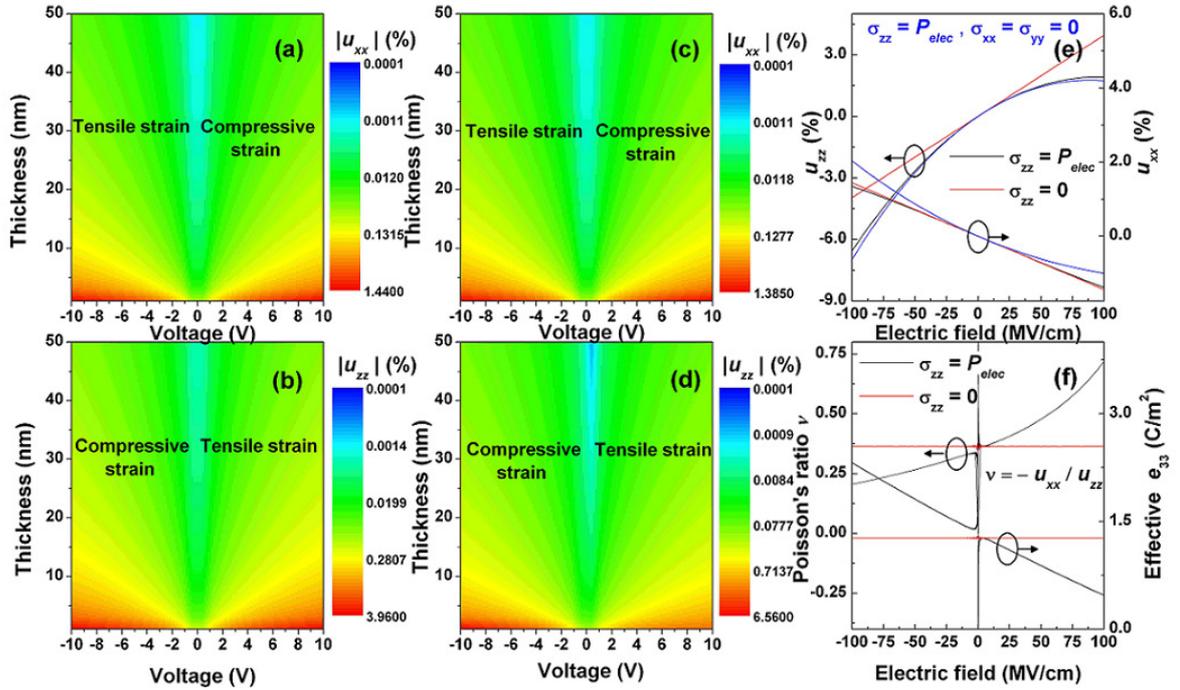

**Figure 2.** Dependence of strain on thickness and applied voltage under constant-**E** conditions for a GaN slab. (a) in-plane strain ($u_{xx}$) without including the electrostatic force correction ($\sigma_{zz}=0$). (b) out-of-plane strain ($u_{zz}$) for $\sigma_{zz}=0$. (c) $u_{xx}$ including the electrostatic force correction ($\sigma_{zz}=P_{elec}$). (d) $u_{zz}$ for $\sigma_{zz}=P_{elec}$. (e) comparison between $u_{xx}$ and $u_{zz}$ as a function of the electric field intensity. (f) dependence of the Poisson's ratio and the effective piezoelectric coefficient $e_{zz}$ on the electric field intensity.

The Eqs. (5)–(7) for $e_{ijk} = 0$ and $D_{0z} = 0$ (i.e., without piezoelectric effect) lead to simple analytical expressions, namely $u_{zz} = -\varepsilon V_a^2/C_{33}d^2$ and $u_{xx} = u_{yy} = 0$. This is a very simple result used to estimate the pure electrostatic force effect. We can observe that $u_{zz}$ is given by the square of the ratio between $V_a$ and $d$, and, consequently, depends quadratically on the electric field. In contrast, the in-plane components are not affected by the electrostatic interaction. Therefore, the electrically-induced Poisson's ratio should be zero. If we add the $\sigma_{xx} = \sigma_{yy} = 0$ boundary conditions, the resulting equations are $u_{zz} = -\varepsilon V_a^2/[C_{33} - 2C_{13}^2/(C_{11} + C_{12})]d^2$ and $u_{xx} = u_{yy} = -C_{13}u_{zz}/C_{11} + C_{12}$. Now, the value of $u_{zz}$ is similarly given by the electric field but with a positive Poisson's ratio. The case of piezoelectric materials is slightly different. Thus, even neglecting the electrostatic interaction between boundary charges, both $u_{xx}$ (also $u_{yy}$) and $u_{zz}$ appear in response to the electric field via the converse piezoelectric effect [see Fig. 2(a)–(b)]. Another remarkable difference is that the compressive or tensile nature of strain depends on the sign of the electric field.

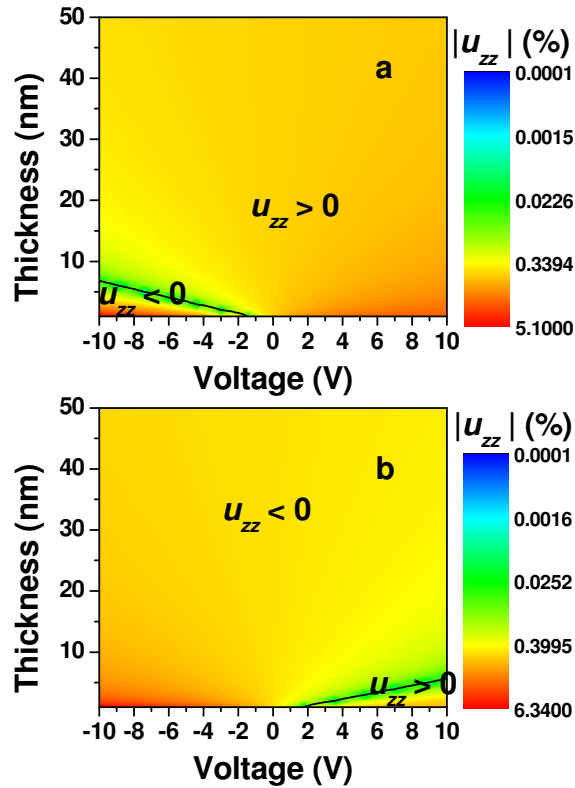

**Figure 3.** Dependence of the out-of-plane strain on thickness and applied voltage under constant-**E** conditions for an in-plane GaN slab constrained to (a) $u_{xx} = u_{yy} = -1\%$ (compressive) and (b) to $u_{xx} = u_{yy} = 1\%$ (tensile).

The combined effects of the electromechanical and electrostatic interactions are shown in Figs. 2(c)–(d) for a GaN slab with a variable thickness in the range of 1-50 nm. These results have been obtained from the calculations using Eqs. (3)–(7) for the constant-**E** condition and applied voltages between −10 and 10 V. We can appreciate that the strain values scale with the intensity of the electric field. Therefore, the strain is larger at the bottom corners of Figs. 2(c)–(d). Note that the dependence on $d$ should be more significant for finite-size electrodes. The results also confirm that the approximation of neglecting the electrostatic interaction is good enough to determine the absolute values of **u** for an electric field up to 260 kV/cm, corresponding to strain values lower than ~0.01%. Moreover, the use of piezoelectric materials increases the effect of the electrostatic force on the overall strain state. Figures 2(e)–(f) show the comparison between the variation of strain with applied voltage including and without the electrostatic interaction [the results adding the $\sigma_{xx} = \sigma_{yy} = 0$ boundary conditions have also been included in Fig. 2(e) for comparison]. There are two significant changes from the qualitative point of view which deserve to be explained in detail. The first one is that the electrostatic interaction breaks the symmetry in magnitude with respect to 0 V. The physical reason for this behaviour is that, in contrast to the piezoelectric effect, the pure electrostatic effect is compressive in nature regardless the sign of the electric field. The asymmetry is more pronounced for increasing values of the electric field. If the intensity of the electric field is low or moderate, the converse piezoelectric effect dominates over the pure electrostatic effect. The effect of the electrostatic interaction becomes indeed comparable only for electric field intensities higher than 10 MV/cm. Nevertheless, the relative variation for the strain including this effect is as high as 12% for $u_{zz}$ at moderate intensities (1 MV/cm), being significantly lower for the in-plane components. The second one is that the Poisson's ratio and the apparent value for $e_{33}$ (piezoelectric coefficient relating $u_{zz}$ to $E_z$), compared to the case without considering the electrostatic interaction, exhibit a nonlinear response due to the dependence of the additional compressive stress on the electric field intensity. We can also observe a resonance for both quantities when either $u_{xx} = 0$ or $u_{zz} = 0$. Note in this regard that the change from compressive to tensile behaviour shown in Figs. 2(a)–(d) is tilted towards positive values of applied voltage for $u_{zz}$ in Fig. 2(c), but not for $u_{xx}$ in Fig. 2(d), where the transition is produced at 0 V regardless the GaN-slab thickness.

The relation between the electrostatic force and the non-linear electrostrictive effects also deserves to be considered. The results described here could suggest that both effects are apparently similar if the electrostatic force and the piezoelectric contributions are considered in a separate way. However, their functional dependences in the fully coupled case are different. The electrostatic force adds a constant term to **u** due to the spontaneous polarization and the resulting values for the strain components are obtained from the solution of the fourth order algebraic equation given by Eq. (5) (second order equation for $\sigma_{xx} = \sigma_{yy} = 0$) instead of from a quadratic law with the electric field. The consequences are significant in practical situations, since the electrostatic force effect could be confused with the electrostrictive effect and vice versa. For example, the value recently reported for the $M_{33}$ electrostrictive coefficient of GaN from theoretical calculations is in the range of $1\times10^{-22}$ m$^2$V$^{-2}$ (note that the experimental confirmation is still lacking) [22]. The curve for $u_{zz}$ in Fig. 2(e) fits to a quadratic expression, leading to first and second order coefficients of $4.26\pm0.01\times10^{-12}$ mV$^{-1}$ and $2.61\pm0.01\times10^{-22}$ m$^2$V$^{-2}$, respectively (higher order effects are negligible in the studied range of electric fields). Therefore, the electrostatic force effect must be considered in the determination of the electrostrictive coefficient of GaN. The quadratic fitting also shows that the piezoelectric coefficient is significantly affected (~10%). In particular, the quadratic correction is smaller than the linear correction under low and moderate electric fields, so that the electrostatic force correction clearly dominates over the electrostrictive contribution in the range of interest. The reason for the linear dependence is found in the linear dependence of $\sigma_{zz}$ with **E** from Eq. (3) through the terms including both **E** and and $D_{0i}$. The mechanical pressure is thus partially equivalent to an extra applied electric field. This result is different to the one obtained in non-piezoelectric materials.

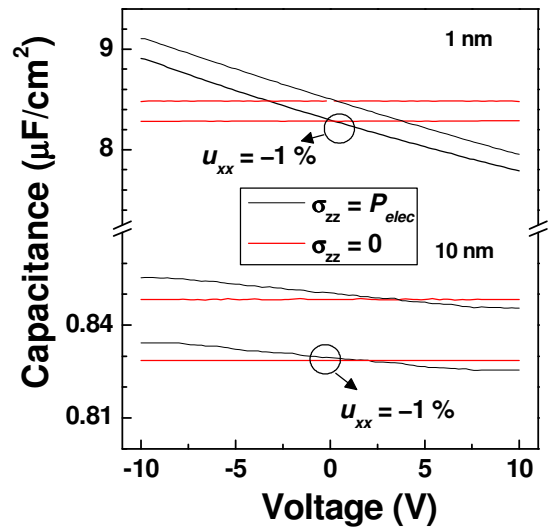

**Figure 4.** Capacitance of a GaN slab as a function of applied voltage at different film thicknesses. The upper panel shows the capacitance for a 1-nm-thick GaN slab, whereas the lower panel shows the capacitance for a 10-nm-thick GaN slab. In both cases, the capacitance is obtained for unconstrained and compressive constrainted ($u_{xx} = u_{yy} = -1\%$) conditions. Black (red) lines correspond to the calculations (without) including the electrostatic force correction, i.e., $\sigma_{zz}=P_{elec}$ ($\sigma_{zz}=0$). The maximum relative variation of capacitance in the displayed range between unconstrained and constrained results is also shown for comparison. Red color online corresponds to gray in the printed version.

Both linear and quadratic contributions due to the electrostatic force depend on the geometry. The simplest case is for circular plates of finite size, where the electrostatic force increases by a factor $1+2d/D$, being $D$ the diameter of the disk. The aspect ratio $d/D$ can be large in nanoscale objects, resulting in remarkable changes in the effective electromechanical coupling as those found in Ref. [13]. These results could also help to explain the variations measured for the piezoelectric coefficients of GaN and related compounds [23−25].

To further investigate the effect of the electrostatic interaction induced by the boundary charge, we considered the case of an isolated clamped slab under both in-plane compressive and tensile constraints. Figure 3 shows the results of calculations for $u_{xx} = u_{yy} = -1\%$ and $u_{xx} = u_{yy} = 1\%$. The strain values determined using the same method as before, and imposing these additional constraints, exhibit a nonzero response at 0 V, even neglecting the electrostatic force between the boundary charges. The change of $u_{zz}$ increases with the electric field, being noticeable for intensities higher than 3 MV/cm, and exceeding the values for the unclamped case only at very large electric fields (>20 MV/cm). This small diference can be attributed to the reduction of the degrees of freedom for the electric enthapy minimization. We can observe that the change of strain as a function of applied voltage can be important in terms of reliability for strained layers close to the critical thickness. For example, the stored elastic energy due to $u_{zz}$ increases 21% for an increase in the amount of strain as high as 10% (e.g., from $u_{zz} = 0.53\%$ to $u_{zz} \approx 0.58\%$). This situation corresponds to $u_{xx} = u_{yy} = -1\%$ and $E_z \approx 1.6$ MV/cm. In that sense, the electrostatic force plays a secondary role compared to the converse piezoelectric effect, except for very large electric fields as abovementioned. The release of the full electrostatic energy into mechanical energy could produce more dramatic effects [26,27].

We can also compare the capacitance of the structure including and without the electrostatic force between electrodes. In both cases, the capacitance reflects the value of the electric permittivity corrected by the electromechanical coupling. The dependence of the capacitance with applied voltage without considering the electrostatic force between electrodes is flat, whereas the effect of the electrostatic force gives rise to an asymmetrical response as a function of voltage (see Fig. 4). The variation of capacitance lies in this case within ±0.8% for the 10-nm-thick GaN slab, increasing to ±7% for the 1-nm-thick GaN slab. On the other hand, the in-plane constraints for $u_{xx}$ and $u_{yy}$ basically shifts the value of capacitance to lower values as shown in Fig. 4. Note that the charge stored depends on both the spontaneous and piezoelectric polarization, but the capacitance is only affected by the voltage-dependent terms, i.e. **E** and **u**.

## 4. Conclusions

In summary, we have theoretically shown that the electrostatic force exerted between the electrodes of a parallel-plate capacitor structure based on a piezoelectric material leads to apparent changes in both the electrostrictive and piezoelectric coefficients. The results of calculations depend on the type of boundary conditions imposed, being the most interesting those for constant-**E** conditions, where the calculations predict the existence of an electrically-induced variation for the Poisson's ratio and the principal piezoelectric coefficient, even at low values of the electric field. The apparent variation in the piezoelectric coefficient is related to the linear dependence of the electrically-induced stress on the electric field through the terms including the spontaneous polarization. The electrostatic force also breaks the symmetry under the applied voltage as observed for strain and capacitance results. These switchable properties are similar to those found in ferroelectric materials [28]. From a quantitative point of view, this interaction affects the accuracy of the measurements at the nanoscale, becoming a first-order effect for extremely intense electric fields and being critical in the quantitative determination of the piezoelectric and electrostrictive response of the material.

The design of multilayer systems with "sacrificial layers" to absorb the electric field and active layers to accommodate the mechanical stress induced by the electrostatic force could give rise to new phenomena. Similarly, the electrostatic force could be also turned to repulsive using a proper structure. In this later case, the value of the piezoelectric coefficient decreases and the behaviour is reversed with increasing the charge density on the electrodes.

**Acknowledgements**

This work was partially supported by the company INSA (Ingeniería y Servicios Aeroespaciales, S.A.).

**Appendix**

The analytical derivation of Eqs. (5)−(7) from the minimization of the electric enthalpy functional uses the Langrange multipliers method. The constraints are given by Eqs. (3)−(4) for constant-**E** conditions. After some algebraic manipulations, the resulting coefficients $\alpha_i$, $\beta_j$ and $\gamma_k$ can be written as

$$\alpha_4 = \frac{4e_{31}e_{33}\beta_2\gamma_1 + 4e_{31}^2\beta_2^2 + e_{33}^2\gamma_1^2}{\varepsilon}, \tag{A1}$$

$$\alpha_3 = 2C_{13}\beta_2\gamma_1 + 4e_{31}\left(E_z + \frac{D_{0z}}{\varepsilon}\right)\beta_2\gamma_1 + \frac{4e_{31}e_{33}(\beta_2\gamma_0 + \beta_1\gamma_1)}{\varepsilon} + \frac{8e_{31}^2\beta_2\beta_1}{\varepsilon} +$$
$$\frac{2e_{33}^2\gamma_1\gamma_0}{\varepsilon} + \left[C_{33} + 2e_{33}\left(E_z + \frac{D_{0z}}{\varepsilon}\right)\right]\gamma_1^2, \tag{A2}$$

$$\alpha_2 = \left[2C_{13} + 4e_{31}\left(E_z + \frac{D_{0z}}{\varepsilon}\right)\right](\beta_2\gamma_0 + \beta_1\gamma_1) + \frac{4e_{31}e_{33}(\beta_1\gamma_0 + \beta_0\gamma_1)}{\varepsilon} + \frac{4e_{31}^2(\beta_1^2 + 2\beta_2\beta_0)}{\varepsilon} +$$

$$2\left[C_{33} + 2e_{33}\left(E_z + \frac{D_{0z}}{\varepsilon}\right)\right]\gamma_1\gamma_0 + \frac{e_{33}^2\gamma_0^2}{\varepsilon} - \left(e_{33}E_z - \varepsilon E_z^2 - \frac{D_{0z}^2}{\varepsilon} - 2D_{0z}E_z\right)\gamma_1^2, \quad \text{(A3)}$$

$$\alpha_1 = \left[2C_{13} + 4e_{31}\left(E_z + \frac{D_{0z}}{\varepsilon}\right)\right](\beta_1\gamma_0 + \beta_0\gamma_1) + \frac{4e_{31}e_{33}\beta_0\gamma_0}{\varepsilon} + \frac{8e_{31}^2\beta_1\beta_0}{\varepsilon} -$$

$$2\left(e_{33}E_z - \varepsilon E_z^2 - \frac{D_{0z}^2}{\varepsilon} - 2D_{0z}E_z\right)\gamma_1\gamma_0 + \left[C_{33} + 2e_{33}\left(E_z + \frac{D_{0z}}{\varepsilon}\right)\right]\gamma_0^2, \quad \text{(A4)}$$

$$\alpha_0 = \left[2C_{13} + 4e_{31}\left(E_z + \frac{D_{0z}}{\varepsilon}\right)\right]\beta_0\gamma_0 + \frac{4e_{31}^2\beta_0^2}{\varepsilon} - \left(e_{33}E_z - \varepsilon E_z^2 - \frac{D_{0z}^2}{\varepsilon} - 2D_{0z}E_z\right)\gamma_0^2, \quad \text{(A5)}$$

$$\beta_2 = \frac{e_{33}}{\varepsilon}\left[2\left(C_{13}e_{33} - \frac{e_{33}^2}{2e_{31}}(C_{11} + C_{12})\right) + 2(C_{13}e_{33} - C_{33}e_{31})\right], \quad \text{(A6)}$$

$$\beta_1 = \left[C_{33} + 2e_{33}\left(\frac{D_{0z}}{\varepsilon} + E_z\right)\right]\left[2C_{13} - (C_{11} + C_{12})\frac{e_{33}}{e_{31}}\right] + 2\left(\frac{D_{0z}}{\varepsilon} + E_z\right)(C_{13}e_{33} - C_{33}e_{31}),$$

$$\text{(A7)}$$

$$\beta_0 = \left[2C_{13} - (C_{11} + C_{12})\frac{e_{33}}{e_{31}}\right]\left[\varepsilon E_z^2 + \frac{D_{0z}^2}{\varepsilon} + E_z(2D_{0z} - e_{33})\right] + E_z(C_{13}e_{33} - C_{33}e_{31}), \quad \text{(A8)}$$

$$\gamma_1 = \frac{2}{\varepsilon}\left[2C_{33}e_{31}^2 - 4C_{13}e_{31}e_{33} + (C_{11} + C_{12})e_{33}^2\right], \quad \text{(A9)}$$

$$\gamma_0 = 2C_{13}\left[C_{13} - 2e_{31}\left(E_z + \frac{D_{0z}}{\varepsilon}\right)\right] - (C_{11} + C_{12})\left(C_{33} - 2e_{33}\left(E_z + \frac{D_{0z}}{\varepsilon}\right)\right) -$$

$$2C_{13}\left[2C_{13} - (C_{11} + C_{12})\frac{e_{33}}{e_{31}}\right], \quad \text{(A10)}$$

where $E_z = V_a/d$. This analysis can be repeated applying additional constraints given by $\sigma_{xx} = \sigma_{yy} = 0$. In this case,

$$u'_{xx} = -\frac{C_{13}}{C_{11} + C_{12}}u'_{zz} + \frac{e_{31}}{C_{11} + C_{12}}E_z, \quad \text{(A11)}$$

and the $\alpha_i$ coefficients of Eq. (5) can be reduced to

$$\alpha'_4 = 0, \quad \text{(A12)}$$

$$\alpha'_3 = 0, \quad \text{(A13)}$$

$$\alpha'_2 = \frac{4e_{31}^2 C_{13}^2}{\varepsilon(C_{11} + C_{12})^2} - \frac{4e_{31}e_{33}C_{13}}{\varepsilon(C_{11} + C_{12})} + \frac{e_{33}^2}{\varepsilon}, \quad \text{(A14)}$$

$$\alpha'_1 = -\frac{2C_{13}^2}{C_{11} + C_{12}} - \frac{8e_{31}^3 C_{13}E_z}{\varepsilon(C_{11} + C_{12})^2} - \frac{4e_{31}C_{13}E_z}{C_{11} + C_{12}} - \frac{4e_{31}C_{13}D_{z0}}{\varepsilon(C_{11} + C_{12})} + \frac{4e_{31}^2 e_{33}E_z}{\varepsilon(C_{11} + C_{12})} +$$

$$C_{33} + 2e_{33}E_z + \frac{2e_{33}D_{z0}}{\varepsilon}, \tag{A15}$$

$$\alpha'_0 = \frac{2C_{13}e_{31}E_z}{C_{11}+C_{12}} + \frac{4e_{31}^4 E_z^2}{\varepsilon(C_{11}+C_{12})^2} + \frac{4e_{31}^2 E_z^2}{(C_{11}+C_{12})} + \frac{4e_{31}^2 D_{z0} E_z}{\varepsilon(C_{11}+C_{12})} - e_{33}E_z + \frac{D_{z0}^2}{\varepsilon} +$$

$$2D_{z0}E_z + \varepsilon E_z^2. \tag{A16}$$